\documentclass[conference]{IEEEtran}
\IEEEoverridecommandlockouts
\usepackage{cite}
\usepackage{amsmath,amssymb,amsfonts}
\usepackage{algorithmic}
\usepackage{graphicx}
\usepackage{textcomp}
\usepackage{xcolor}
\usepackage{comment}
\usepackage{lipsum}
\usepackage{enumitem}

\usepackage{booktabs}
\usepackage{colortbl}
\usepackage[table]{xcolor}
\usepackage{array}
\usepackage{caption}
\usepackage{makecell}
\usepackage{ctable}
\usepackage{footnote}

\usepackage{booktabs}
\usepackage{threeparttable}

\usepackage{subcaption}
\usepackage{dblfloatfix}

\usepackage{multirow}      

\usepackage{float}
\usepackage{nicefrac}
\usepackage{pifont}
\newcommand{\ytick}{{\ding{51}}}
\newcommand{\ntick}{{\ding{55}}}

\def\BibTeX{{\rm B\kern-.05em{\sc i\kern-.025em b}\kern-.08em
    T\kern-.1667em\lower.7ex\hbox{E}\kern-.125emX}}

\newcommand{\timescompact}{{\mkern-2mu\times\mkern-2mu}}
\newcommand{\simcompact}{{\mkern-1mu\sim\mkern-1mu}}
\newcommand{\approxcompact}{{\mkern-1mu\approx\mkern-1mu}}
\newcommand{\dash}{\mbox{--}}

\definecolor{ipmagenta}{HTML}{a82fa8}
\newcommand{\IP}[1]{\textcolor{ipmagenta}{#1}}
\definecolor{ipred}{HTML}{FF0000}

\definecolor{dmblue}{HTML}{0b03fc}
\newcommand{\DM}[1]{\textcolor{dmblue}{#1}}

\begin{document}

\title{Optimizing GEMM for Energy and Performance on Versal ACAP Architectures}

\author{\IEEEauthorblockN{Ilias Papalamprou, Dimosthenis Masouros, Ioannis Loudaros, Francky Catthoor, Dimitrios Soudris}
\IEEEauthorblockA{\textit{National Technical University of Athens}}
}

\maketitle

\begin{abstract}
General Matrix Multiplication (GEMM) is a fundamental operation in many scientific workloads, signal processing and particularly deep learning. 
It is often a bottleneck for performance and energy efficiency, especially in edge environments with tight resource and power constraints.
AMD’s Versal ACAP offers heterogeneous components (AIEs, PL, PS) that can address these challenges, but mapping GEMM across them is complex, with prior works largely overlooking energy-performance tradeoffs. 
In this paper, we propose an automated framework for Versal ACAP that generates GEMM mappings optimized for either performance or energy efficiency.
Unlike prior analytical approaches, our method leverages a Machine Learning (ML) model, trained on $\approx$6000 on-board experiments of different GEMM mappings, to guide Design Space Exploration, yielding more efficient designs.
Evaluation on the Versal VCK190 shows geomean improvements of 1.23$\timescompact$ (up to 2.5$\timescompact$) in throughput and 1.25$\timescompact$ (up to 2.7$\timescompact$) in energy efficiency over state-of-the-art frameworks.

\end{abstract}

\section{Introduction}
General Matrix Multiplication (GEMM) is a fundamental  kernel in scientific computing, signal processing, and especially deep learning (DL), where it can account for up to $99\%$ of total FLOPs~\cite{kim2023full}, in classical Multilayer Perceptrons (MLPs)~\cite{popescu2009multilayer} and Large Language Models (LLMs)~\cite{brown2020language}. 
As the primary computational bottleneck, GEMM efficiency directly determines both performance and energy consumption.
This challenge is further amplified in embedded systems deployed in \emph{edge computing} environments~\cite{hua2023edge,murshed2021machine}, where strict power budgets, limited cooling, and constrained on-chip resources prevent sustaining peak throughput at the cost of excessive power draw~\cite{tu2023unveiling,gutierrez2021automatic,maity2021thermal}.
In these settings, energy efficiency (e.g., FLOPs per Watt) becomes the decisive metric, requiring co-optimization of performance and power to sustain high efficiency under resource and power constraints~\cite{sahin2015impacts}.
AMD’s Versal Adaptive Compute Acceleration Platform (ACAP)~\cite{gaide2019xilinx} is a compelling target for achieving both high-performance and energy efficient execution.
It's architecture integrates three distinct components: 
\emph{i)}~AI Engines (AIEs) with VLIW, SIMD and local scratchpad memory, \emph{ii)}~programmable logic (PL) for custom functions and flexible datapaths, and \emph{iii)}~embedded CPU as the processing system (PS) for orchestration.
These components are interconnected via a Network-on-Chip (NoC), enabling fine-grained customization of compute-memory pipelines.
This heterogeneity offers substantial opportunities for energy-efficient GEMM acceleration, but also in the same time it presents three main challenges:

\begin{enumerate}
    \item \textbf{Complex hardware mapping:}
    Executing GEMM workloads across heterogeneous components (AIEs, PL and PS) produces a vast design space with thousands of possible mapping options ($>$6000 for typical GEMM operations), making manual exploration impractical.
    Automated frameworks are essential to perform design space exploration (DSE) and identify efficient mappings.
    \item \textbf{Restricted on-chip memory:} Limited local memory in the AIEs and low DDR bandwidth further complicate workload mapping.
    \emph{Tiling strategies} are often employed, which split computation into smaller tiles, improving data reuse and alleviating bandwidth pressure (see Sec.~\ref{sec:background}).
    Yet, finding optimal tiling factors is non-trivial, as they directly determine the data transferred to and from DDR and the number of AIEs that can be effectively utilized, thereby affecting performance and power consumption.
    \item \textbf{Unreliable power modeling:}
    Diverse operations across heterogeneous components, such as vectorized instructions, NoC transfers and memory accesses in AIEs, PL and DDR, make system power consumption challenging to predict.
    Vendor-provided tools often diverge substantially from real hardware measurements~\cite{lin2022powergear,wang2025reconfigurable}, hindering energy-aware optimization. Accurate power modeling is required to efficiently guide the DSE process.
    
\end{enumerate}

\begin{figure}[t]
    \centering
    \begin{subfigure}{0.73\linewidth}
        \centering
        \includegraphics[width=\linewidth]{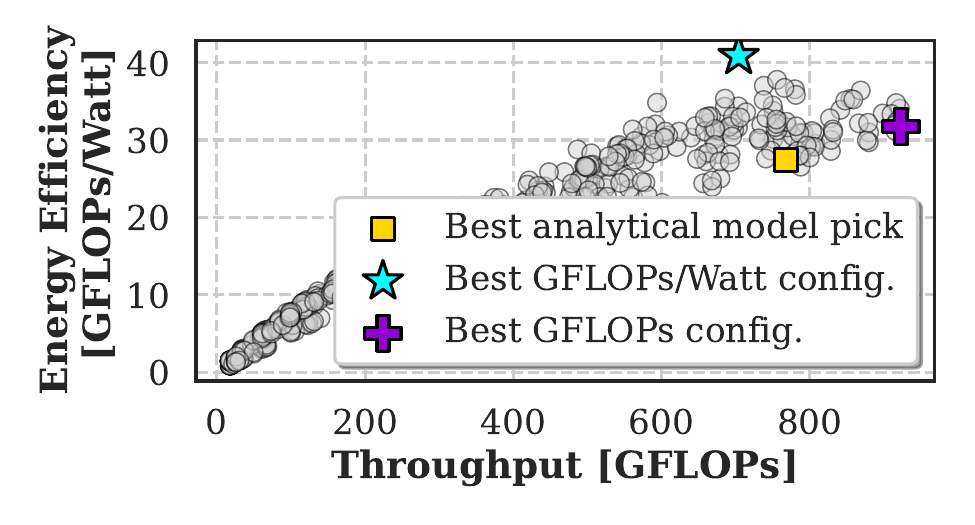}
        \caption{Throughput and energy efficiency.}
        \label{fig:intro_energy_throughput}
    \end{subfigure}
    \hfill
    \begin{subfigure}{0.25\linewidth}
        \centering
        \includegraphics[width=\linewidth]{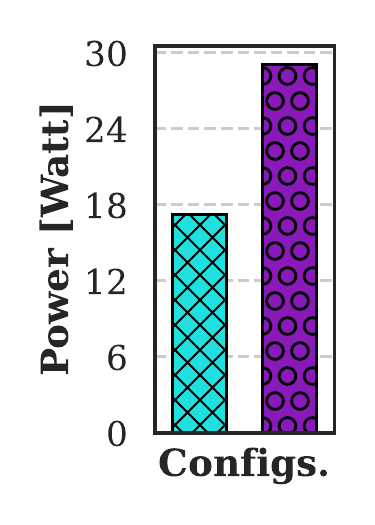}
        \caption{Power of best designs.}
        \label{fig:intro_power}
    \end{subfigure}
    \caption{Impact of tiling on GEMM performance and power.}
    \label{fig:intro_overall}
\end{figure}

Several prior works explore GEMM acceleration on Versal ACAP.
However, they focus on maximizing performance either through high throughput AIE kernels~\cite{zhuang2023high}, multiple units for diverse GEMM workloads~\cite{zhuang2023charm,zhuang2024charm,wang2025reconfigurable} or by improving data reuse on the AIEs~\cite{taka2023maxeva,taka2024efficient,mhatre2025gama}.
To guide DSE, they typically rely on analytical models to estimate performance and select tiling parameters, which in turn determine AIE allocation and buffer sizes.
While effective for throughput optimization, this strategy introduces two key limitations.
First, \emph{power is largely overlooked}.
An implicit assumption in these works is that maximizing throughput naturally yields optimal energy efficiency, but this is not always the case.
Fig.~\ref{fig:intro_energy_throughput} shows the throughput and energy efficiency of different tiling configurations for a GEMM workload.
We observe that the highest-throughput design (purple cross) is 22.4$\%$ less energy-efficient than the most energy-efficient one (teal star). 
This is caused by higher AIE utilization, which results in a $\approx$11~Watt increase in power consumption (Fig. 1b).
Second, \emph{analytical models are prone to inaccuracies}.
While the DSE process can tolerate small estimation errors when only relative differences matter, large deviations risk overlooking optimal mappings.
As shown in Fig.~\ref{fig:intro_energy_throughput}, analytical-model based mappings (yellow square) can reduce throughput by 17\% compared to the actual highest-throughput design.
Moreover, such models also neglect configurations that are superior not only in performance but also in energy efficiency (gray circles).

To address the above limitations, we introduce an automatic framework that optimizes GEMM mappings on Versal ACAP architectures.
Given the GEMM dimensions and a target objective (throughput or energy efficiency) our framework performs DSE driven by a machine learning (ML) model.
Trained on a custom dataset from thousands of on-board experiments, the model predicts latency, power, and resource utilization with high accuracy.
This enables efficient exploration of the design space and identification of Pareto-optimal mappings along the energy-performance frontier, yielding implementations tuned for either high throughput or energy-efficient operation.
To enhance robustness across unseen workloads, custom-crafted features are integrated into the ML model. 
To the best of our knowledge, this is the first work to optimize GEMM mappings for both performance and energy efficiency on Versal ACAP.
The novel contributions of this work are as follows:
\begin{itemize}[leftmargin=*]
    \item \textbf{We create an open-source dataset} of $\approx$6000 different GEMM mappings with experiments on Versal VCK190\footnote{Will be released open source upon acceptance.}.
    \item \textbf{We develop an ML-model} enhanced with custom-crafted features, that predicts throughput, energy and allocated resources of GEMM designs, achieving 51$\%$ higher accuracy compared to analytical-based modeling approaches~\cite{zhuang2025aries}.
    \item \textbf{We introduce an automatic framework} that performs DSE to identify Pareto-optimal hardware mappings for a given workload, enabling configurations optimized for either throughput or energy efficiency.
\end{itemize}

\noindent We evaluate our framework on diverse GEMM workloads and compare it against state-of-the-art approaches~\cite{zhuang2023charm,zhuang2025aries} on Versal ACAP and NVIDIA embedded GPUs.
Our experimental results showcase geomean improvements of 1.23$\times$ and up to 2.52$\times$ in throughput, and geomean 1.25$\times$ and up to 2.69$\times$ in energy efficiency over the state-of-the-art.

\section{Related Work}

\begin{table}
\centering
\caption{Comparison with Related Work}
\label{tab:related_dse}
\renewcommand{\arraystretch}{1.2}
\resizebox{\linewidth}{!}{
\begin{tabular}{cccccc}
\specialrule{.2em}{.0em}{0em}
\rowcolor[HTML]{c7c3c3}
\shortstack{\textbf{Prior} \\ \textbf{Works}}
 & \shortstack{\textbf{Performance} \\ \textbf{Modeling}} 
 & \shortstack{\textbf{Considered} \\ \textbf{Components}} 
 & \shortstack{\textbf{Predicted} \\ \textbf{Metrics$^\dag$}}
 & \shortstack{\textbf{Energy-Perf.} \\ \textbf{Tuning}} \\
 \specialrule{.2em}{.0em}{0em}
\textbf{\cite{dai2024g2pm}} & ML-based & AIE & $\mathcal{T}$ & \ntick \\
\hline
\rowcolor[HTML]{EFEFEF}
\textbf{\cite{zhuang2023charm,zhuang2024charm,zhuang2023high,zhuang2025aries,dong2024eq,zhuang2024ssr,wang2025reconfigurable}} & Analytical & AIE, PL, DDR & $\mathcal{T},\mathcal{R}$ & \ntick \\
\hline
\textbf{\cite{taka2023maxeva,taka2024efficient,deng2024ama,mhatre2025gama}} & Analytical & AIE, PL & $\mathcal{T},\mathcal{R}$ & \ntick \\
\hline
\rowcolor[HTML]{CCFFCC}
\textbf{This Work} &  ML-based & AIE, PL, DDR & $\mathcal{T},\mathcal{P},\mathcal{R}$ & \ytick \\
\specialrule{.2em}{.0em}{0em}
\end{tabular}
}
\begin{tablenotes} \footnotesize 
        \item $^\dag$ $\mathcal{T}$: Throughput, $\mathcal{P}$: Power, $\mathcal{R}$: Resources
    \end{tablenotes}
\label{tab:related}
\end{table}

Several works have focused on accelerating GEMM operations on Versal ACAP, to support DL applications.
These include designing efficient AIE-based kernels, architectural optimizations for multiple workload mapping and methodologies that simplify hardware development.
Works~\cite{zhuang2023charm,zhuang2024charm,zhuang2023high} implement efficient matrix multiplication kernels on AIEs, using pipelined designs and optimized PL-AIE communication interfaces.
In~\cite{wang2025reconfigurable} authors propose an ISA abstraction layer on top of the Versal ACAP to orchestrate the available resources, providing rapid hardware reprogramming.
Additionally, works~\cite{zhuang2024ssr,dong2024eq} deploy Vision Transformer (ViT) models on Versal by storing weights in AIE local memories.

Authors in~\cite{taka2023maxeva,taka2024efficient} propose a hybrid GEMM architecture combining matrix multiplication kernels with adder trees, while~\cite{mhatre2025gama} introduces custom buffer placement to maximize AIE memory utilization.
In~\cite{deng2024ama} a hierarchical performance analysis model is introduced to improve GEMM accelerator efficiency.
However, the results in these works are based solely on simulations, without validation on real hardware.

Works~\cite{zhuang2025aries,li2022compiler,dai2024g2pm,mhatre2025performance} propose methodologies to facilitate accelerator development for Versal ACAP.
\cite{dai2024g2pm}~introduces a hierarchical graph-based modeling technique, for faster performance estimation of AIE designs compared to vendor tools.
Nevertheless, model's training is conducted with simulation data and focusing only on the AIEs, overlooking the energy of the deployed designs.
Authors in~\cite{zhuang2025aries} address the complexity of developing accelerators, by proposing an abstraction framework that leverages MLIR to automatically generate low-level code for AIEs, PL and host code from high-level languages, such as Python.
Lastly, authors in~\cite{mhatre2025performance} examine the mapping of GEMM operations on Versal ACAP architecture, revealing insights for the execution and mapping of such workloads.
Their analysis  is mainly based on analytical equations, with limited number of hardware experiments.

Table~\ref{tab:related} summarizes the comparison between prior work and our approach.
Our work:
\emph{i)} uses an ML-model for predicting $\mathcal{T}$, $\mathcal{P}$, and $\mathcal{R}$, trained from real data.
\emph{ii)} Supports tuning GEMM mapping for energy-efficiency or performance.
\section{Background \& Motivation}

\subsection{GEMM Implementation on Versal ACAP}
\label{sec:background}

\begin{figure}[t]
    \centering
    \includegraphics[width=0.8\linewidth]{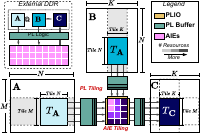}
    \caption{Tiled GEMM operation on Versal ACAP.
    }
    \label{fig:background_mapping}
\end{figure}

AMD's Versal ACAP~\cite{gaide2019xilinx} is a heterogeneous SoC that, unlike traditional FPGAs, integrates AIEs alongside the PL and PS.
The AIEs are arranged in a 2D array (e.g., 50$\times$8 on the VCK190 evaluation board) and connected via AXI interfaces.
The most effective method of mapping GEMM on Versal ACAP is to partition the matrix multiplication operation into \emph{tiles}, to maximize performance by exploiting data reuse~\cite{zhuang2023charm,zhuang2024charm,taka2023maxeva,wang2025reconfigurable,zhuang2025aries}.
As illustrated in Fig.~\ref{fig:background_mapping}, input matrices $\mathbf{A},\mathbf{B}$ initially reside in DDR memory. 
During execution, tiles $T_{\mathbf{A}}$ and $T_{\mathbf{B}}$ are transferred to the PL, where they are buffered before being forwarded to multiple AIEs for computation.
Each AIE executes optimized kernels on a fixed tile size, achieving up to $\approx$90\% of peak performance.
The PL then collects the partial results, forms the output tile $T_\mathbf{C}$, and writes it back to DDR. 
This process repeats until all tiles are computed.
Tiling parameters determine: \emph{i)}~workload parallelization across AIEs and \emph{ii)} the size of PL data reuse buffers along each matrix dimension $d\in\{M,N,K\}$.
These choices directly affect how compute and memory resources are allocated. Smaller allocations reduce power but limit throughput, since fewer AIEs are active and less reuse is achieved.
Larger allocations increase throughput by exploiting more parallelism and data reuse, but they also raise power consumption and place greater demand on memory bandwidth.
\textit{Manual tuning of tiling parameters on the Versal ACAP is complex, as it demands co-optimizing data movement across DDR, PL, and AIEs, within a large design space created by fully custom PL buffers and flexible AIE allocation.}
Therefore, identifying the right tiling configuration requires DSE, with prior work relying on analytical models to estimate performance and select the best candidate configurations~\cite{zhuang2023charm,zhuang2024charm,zhuang2025aries,taka2023maxeva}.

\subsection{Motivation \& Opportunities for Energy Efficient GEMM}
\label{sec:insights}
Next, we study the impact of tiling on power, performance and energy efficiency.
Our observations are drawn from extensive experiments on the VCK190 board, covering $\approx$6000 hardware designs collected over more than 40 days.

\subsubsection{Impact of tiling on power consumption}
\label{sec:insights:power}
\begin{figure}
    \centering
    \includegraphics[width=1.0\linewidth]{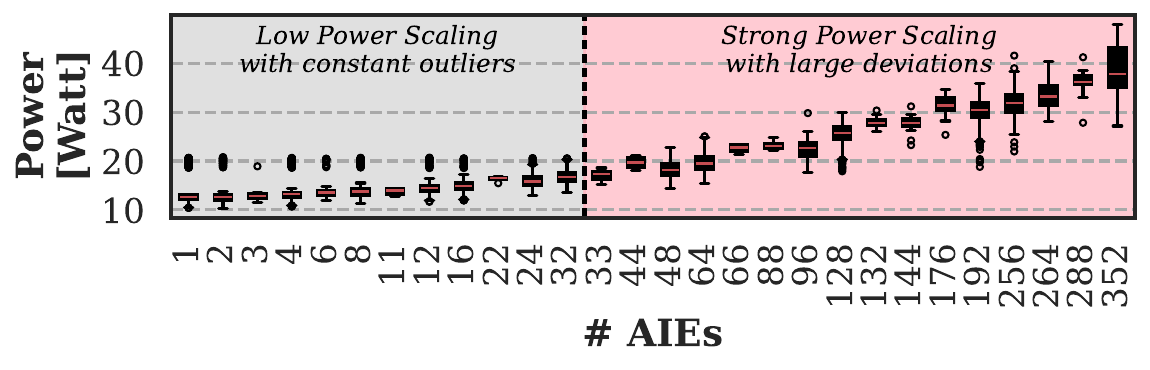}
    \caption{
    System power for varying AIE utilization.
    }
    \label{fig:power_aie_scaling}
\end{figure}

First, we investigate how tiling affects power consumption.
We analyze GEMM workloads with varying $M\timescompact N\timescompact K$ dimensions and tiling configurations, which affect both AIE utilization and PL data-reuse buffer sizes.
Fig.~\ref{fig:power_aie_scaling} shows how system's power consumption changes with the number of active AIEs.
Note that a given AIE utilization can correspond to multiple mappings, since different tiling choices may allocate different amounts of PL data-reuse buffers for the same number of AIEs.
For example, a GEMM can be mapped on 256 AIEs (AIE tiling $[8,8,4]$), with PL tiling $[1,1,1]$ for minimum data reuse or tiling $[4,8,1]$ for much higher data reuse, resulting in 33$\times$ increase in PL memory.
Based on the median values (red lines), power increases with \#AIEs but non-uniformly.
From 1 to 32 AIEs, power rises gradually from 12 to 18~Watt, with nearly constant outliers around 10-20~Watt.
Beyond 32 AIEs, power grows more steeply as AIE dynamic power dominates, with medians from 19-38~Watt.
Outliers show large variation up to 20~Watt, with peak at $\approx$49~Watt.
This span is mainly caused by different PL buffer tiling, which alters data reuse and DDR transfers.
Consequently, some workloads with more AIEs can use less power than others with fewer AIEs.

\begin{figure}[t]
    \centering
    \includegraphics[width=1.0\linewidth]{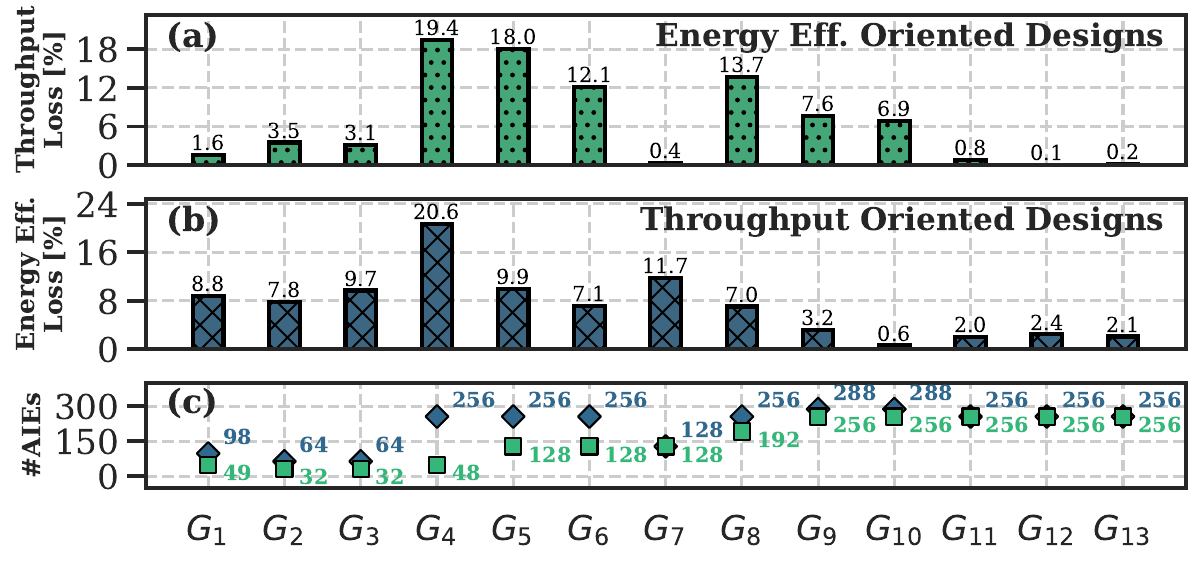}
    \caption{Trade-offs between energy- and throughput-oriented mappings: (a) throughput loss of energy-oriented, (b) energy efficiency loss of throughput-oriented, (c) AIE utilization.}
    \label{fig:performance_energy_tradeoff}
\end{figure}

\subsubsection{Energy-Performance Trade-offs}
We also analyze trade-offs between the most energy-efficient and the highest-throughput designs.
Fig.~\ref{fig:performance_energy_tradeoff}a-c report: \textit{a)} throughput loss of energy-oriented designs (\%); \textit{b)} energy efficiency loss of throughput-oriented designs (\%) and \textit{c)} AIE utilization, across GEMM workloads ($G_n$), sorted by increasing FLOPs.
The observed trade-offs depend strongly on workload size.
For low-FLOP workloads ($G_1\dash G_3$), energy-oriented designs yield large efficiency gains with only 1.6–3.1$\%$ throughput loss, while throughput-oriented ones incur 7.8–9.7$\%$ efficiency loss.
Fig.~\ref{fig:performance_energy_tradeoff}c shows that energy-oriented designs use roughly half the AIEs. 
These gains arise from the small GEMM dimensions, which maximize PL data reuse and maintain efficiency even with fewer AIEs.
Medium-FLOP workloads ($G_4\dash G_{10}$) exhibit the largest trade-offs, with energy-oriented designs losing 0.4–19.4$\%$ throughput and throughput-oriented designs losing 3.2–20.6$\%$ in energy efficiency. 
Here, AIE utilization differs most, with energy-oriented mappings using 1.12–5.3$\times$ fewer AIEs, as the matrices dimensions allow multiple tiling options, exposing trade-offs between the optimization goals.
For high-FLOP workloads ($G_{11}\dash G_{13}$), trade-offs are negligible (0.1–2.1$\%$), as the most energy-efficient designs closely match the highest-throughput ones.
Both designs share the same number of AIEs, indicating that reducing \#AIEs at high FLOPs imposes a greater latency penalty than any power savings. 

\section{Automated Energy-Efficient and High-Performance Gemm Mapping on Versal Acap}
\begin{figure}[t!]
    \centering
    \includegraphics[width=1.0\linewidth]{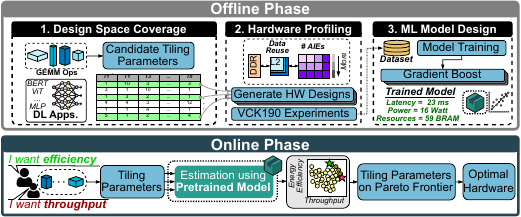}
    \caption{Proposed framework architecture.}
    \label{fig:framework}
\end{figure}

Our framework addresses the challenge of GEMM mapping on Versal ACAP by generating Pareto-optimal designs along the performance–energy frontier. 
Given the GEMM dimensions and a target objective (throughput or energy efficiency), it performs DSE guided by an ML model.
As shown in Fig.~\ref{fig:framework}, the workflow consists of an \emph{offline} phase, where the ML model is developed and trained to predict latency, power, and resource utilization, and an \emph{online} phase, where the pretrained model is used to efficiently identify optimal mappings.

\subsection{Offline Phase}
The \emph{Offline Phase} builds the ML model that drives the DSE.
GEMM operations are collected from popular DL applications and an analytical model is used to select tiling configurations spanning best, worst, and intermediate cases.
These designs are executed on-board to measure performance and power consumption.
The resulting dataset trains a global model that can generalize to unseen GEMM sizes.


\subsubsection{Design Space Coverage}
To generate a comprehensive dataset, we extract multiple GEMM operations from popular DL applications, including NCF, MLP, ViT, BERT as in~\cite{zhuang2023charm,zhuang2024charm,wang2025reconfigurable,zhuang2023automm}. 
These applications contain matrices of varying dimensions $M \timescompact N \timescompact K$, resulting in a diverse set of GEMM workloads $G_n$.
We explore different combinations of tiling parameters, that define \emph{a)}~the number of allocated AIEs, thus, varying the workload parallelization ($P_d$), and \emph{b)}~the architecture of the data reuse buffers in the PL ($B_d$), where $d \in \{M,N,K\}$.
In each AIE, a $32 \timescompact 32 \timescompact 32$ workload is processed, as it offers high performance compared to different tiling configurations~\cite{mhatre2025performance,zhuang2023charm,zhuang2024charm}.

For each GEMM workload $G_n$, we define $C(G_n)$ as the set of candidate tiling parameters that evenly partition the dimensions $d$ of $G_n$.
Since it is infeasible to generate and evaluate hardware designs for all possible mappings ($\approx$6000 per workload), we adopt a sampling approach to efficiently capture the design space.
To do so, we use analytical equations from~\cite{zhuang2025aries} to select a subset of candidate configurations $S(G_n) \subset C(G_n)$ that includes top-performing, worst-performing, and randomly chosen intermediate designs.
This ensures that each GEMM workload is mapped across the full range of AIE allocations, while also exploring different PL buffer sizes, producing a representative dataset that enables the model to generalize. 
To account for inaccuracies of the analytical model, we apply relaxed resource constraints, preventing potentially optimal configurations from being excluded.

\subsubsection{On-board Hardware Profiling} 
Based on the candidate tiling configurations $S(G_n)$, we generate the corresponding hardware designs and evaluate them on-board.
For each configuration $S(G_n)$, we generate the AIE and PL bitstream along with the host executable, using the ARIES framework~\cite{zhuang2025aries}. 
We then retain only successful builds and collect on-board latency and power measurements directly.
In total, the dataset comprises $\approx$6000 hardware designs across 18 GEMM workloads.
Inline with prior work~\cite{zhuang2023charm}, we focus on FP32 datatype.
While newer formats (e.g., bfloat16) are common in DL inference~\cite{kalamkar2019study,burgess2019bfloat16,jeong2022tensorrt}, they are not supported on the VCK190.

\subsubsection{ML Model Design}
\label{sec:impl:training}
Last, we train and validate the ML model that drives the DSE process for finding optimal GEMM mappings.
Since the input features of each workload are bounded by the range of tiling parameters, we adopt Gradient Boosted Decision Trees~\cite{chen2016xgboost}, which are well-suited for accurate prediction on bounded datasets~\cite{malistov2019gradient}. To improve accuracy, we train separate models for latency ($\mathcal{L}$) and power ($\mathcal{P}$), along with a multi-output model for PL resource utilization ($\mathcal{R}$).

We consider 17 \emph{model features}, organized into two categories: Set-I of fundamental parameters and Set-II of complementary features.
Set-I includes features directly obtained from the GEMM workload $G_n$, namely the GEMM dimensions $d \in \{M,N,K\}$ and the candidate tiling parameters for AIEs and PL data-reuse buffers ($P_d$ and $B_d$).
Set-II includes custom-crafted features that capture interactions between the workload and the hardware configuration, such as metrics derived from AIE allocation and various workload-to-tiling ratios.

Specifically, we compute the allocated AIEs as $N_{\text{AIE}}= P_M \cdot P_N \cdot P_K$ and the computational load over the allocated AIEs as $\rho=\nicefrac{\text{FLOP}}{N_\text{AIE}}$, where FLOP denotes the total floating-point operations of $G_n$.
Notably, $\rho$ is strongly correlated with the GEMM execution time (Pearson Correlation $r=0.81$), as the AIEs are the main processing elements and largely determine performance.
Set-II also includes ratios ($R_{P_i}$ and $R_{B_i}$) between each workload dimension $d$ and each tiling factor, allowing the model to capture how tiling behavior scales across different workload dimensions, and thus improve generalization on unseen cases.
The model features $\Phi$ are:
\begin{equation*}
\Phi = 
\Bigl\{
    \underbrace{d, P_d, R_d}_{\text{Set-I}}, \;
    \underbrace{N_{\text{AIE}}, \rho, R_{P_d}, R_{B_d}}_{\text{Set-II}}, \;
    \;\Big|\; d \in \{M,N,K\} 
\Bigr\}
\label{eq:features}
\end{equation*}

\begin{figure}[t]
    \centering
    \begin{minipage}{0.49\linewidth}
        \centering
        \includegraphics[width=\linewidth]{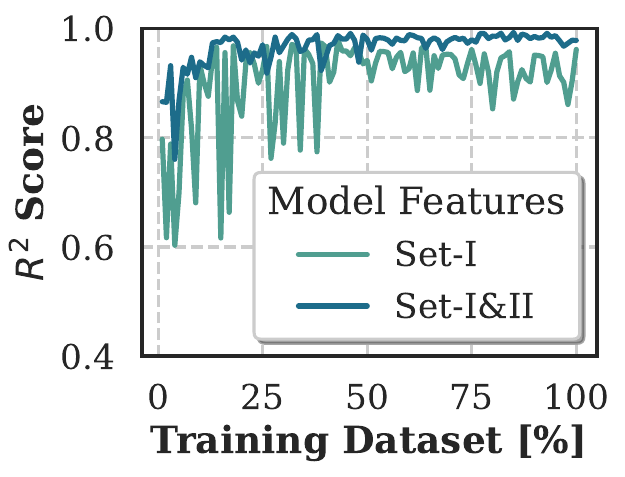}
        \caption{$R^2$ score for variable samples in the training dataset.}
        \label{fig:r2_progress}
    \end{minipage}\hfill
    \begin{minipage}{0.43\linewidth}
        \centering
        \includegraphics[width=\linewidth]{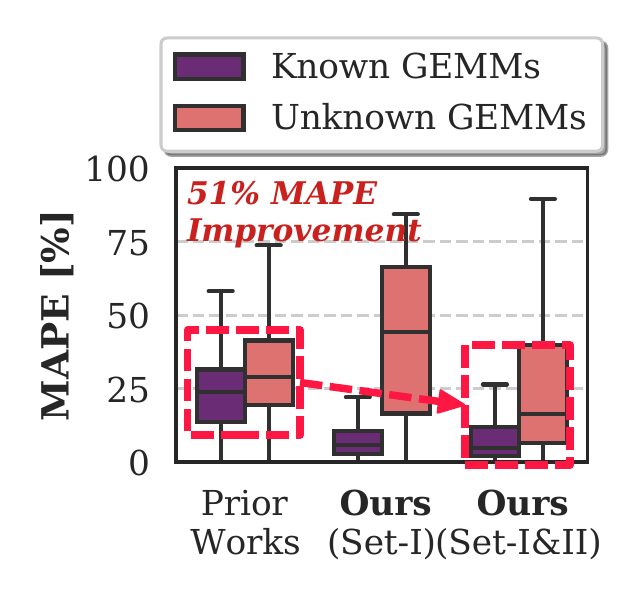}
        \caption{Prediction error of proposed ML vs. analytical model
        }
        \label{fig:model_mape}
    \end{minipage}
\end{figure}

\noindent\textbf{\underline{Model Training \& Validation:}}
\label{sec:impl:model_eval}
We train the models using an 80/20 train-test split with 5-fold cross-validation and optimize hyperparameters via Bayesian optimization with Optuna~\cite{akiba2019optuna}. 
For the $\mathcal{L}$ model, we apply a logarithmic transformation on latency to reduce variance. 
Fig.~\ref{fig:r2_progress} shows the $R^2$ score of $\mathcal{L}$ under different training set sizes, illustrating how prediction accuracy improves as more hardware designs become available.
Using both Set-I\ and Set-II, the model achieves a high $R^2$ of 0.986 even with $\approx$30$\%$ of the data, demonstrating accurate predictions on sparse datasets. 
For both sets, $R^2$ scores evolve smoothly, with Set-I showing higher variance.
Fig.~\ref{fig:model_mape} also shows the Mean Absolute Percentage Error (MAPE, $\%$) of our $\mathcal{L}$ model versus prior analytical models~\cite{zhuang2025aries} for \emph{(a)} known and \emph{(b)} unknown GEMM workloads. 
Analytical models achieve a median MAPE of 26.67$\%$, generally showing higher accuracy for square GEMM shapes but lower accuracy for more complex GEMM shapes.
Our model achieves 34.16$\%$ with Set-I and 13.09$\%$ with Set-I\&II, improving accuracy by 50.9$\%$.
For known workloads, both perform similarly (5.82$\%$ vs. 4.77$\%$, $\approx$8$\%$ improvement). 
For unknown workloads, adding Set-II reduces MAPE from 44.19$\%$ to 16.52$\%$ (62.62$\%$ improvement). 
Similar behavior is observed for $\mathcal{P}$ and $\mathcal{R}$ models, with low MAPE of 7.05$\%$ and 6.046$\%$. 

\begin{table}[b]
\centering
\caption{Evaluation Setup}
\resizebox{\linewidth}{!}{ %
\begin{tabular}{ccccc}
\specialrule{.2em}{.0em}{0em}
\rowcolor[HTML]{c7c3c3}
& \makecell{\textbf{GPU I} \\ AGX Xavier} & \makecell{\textbf{GPU II} \\ Xavier NX} & \makecell{\textbf{GPU III} \\ AGX Orin} & \makecell{\textbf{Versal ACAP} \\ VCK190} \\
\specialrule{.2em}{.0em}{0em}
\textbf{Computing Resources} & 64 TC & 48 TC & 64 TC & AIEs \& PL$^\dag$ \\
\hline
\rowcolor[HTML]{EFEFEF}
\textbf{Technology [nm]} & 12 & 12 & 8 & 7 \\
\hline
\textbf{Peak Perf. [GFLOPS]} & 1410 & 844.8 & 5325 & 8000 \\
\hline
\rowcolor[HTML]{EFEFEF}
\textbf{Memory BW [GB/s]} & 136.5 & 59.71 & 204.8 & 25.6 \\
\specialrule{.2em}{.0em}{0em}
\end{tabular}
}
\begin{tablenotes} 
    {\scriptsize 
    \item $^\dag$ 400 AIEs, PL: 963 BRAM, 463 URAM, 900K LUTs, 1.8M FFs, 1.9K DSPs
    }
\end{tablenotes}
\label{table:exp_setup}
\end{table}

\subsection{Online Phase}
The goal of the \emph{online phase} is to generate the Pareto-optimal mapping for a given GEMM workload based on the optimization objective (high throughput or energy efficiency). 
The process begins with the user specifying the GEMM workload and optimization goal. 
As in the \emph{offline phase}, our framework then enumerates all possible tiling configurations $T(P_i,B_i)$.
Furthermore, using the workload dimensions and tiling parameters ($P_i$, $B_i$), features from Set-II are computed.

\subsubsection*{ML-driven DSE}
For each possible tiling parameter combination $T(P_i,B_i)$, using the pretrained models from the \emph{offline phase} our framework predicts the target metrics $\{\mathcal{L},\mathcal{P},\mathcal{R}\}$.
Afterwards, the tiling parameter combinations that fit in the available PL resources are filtered, ensuring valid hardware designs.
This leads to a set with candidate GEMM mappings, with predictions of throughput, energy efficiency and detailed resource utilization for each one.
From the generated Pareto front, we can select the best configuration based on the user requirements, whether the prioritization leads towards energy efficiency or high performance.

\section{Experimental Setup and Evaluation}
\label{sec:eval:exp_setup}
The devices used in all experiments are listed in Table~\ref{table:exp_setup}, along with their specifications. 
For our experiments targeting Versal ACAP, we use the VCK190 evaluation board that features the XCVC1902 device.
The AIEs operate at 1.25GHz, while the kernels deployed on the PL operate at 230 MHz.
The designs are developed using AMD Vitis 2023.2.
To measure power consumption of VCK190, each workload is executed for 60 seconds, during which power data is collected via BEAM tool running on Versal's System Controller (SC). 
In the listed experimental results, the total power of the evaluation board is reported.
To compare performance with other platforms, we also evaluate our framework with equivalent implementations on three Nvidia Jetson embedded GPUs.
The GPU workloads are implemented in PyTorch and executed using the latest version of CUDA.
The total power consumption of Jetson boards is measured with similar methodology as for the VCK190, while the power samples are acquired using Tegrastats.

\begin{figure}[t]
    \centering
    \includegraphics[width=1.0\linewidth]{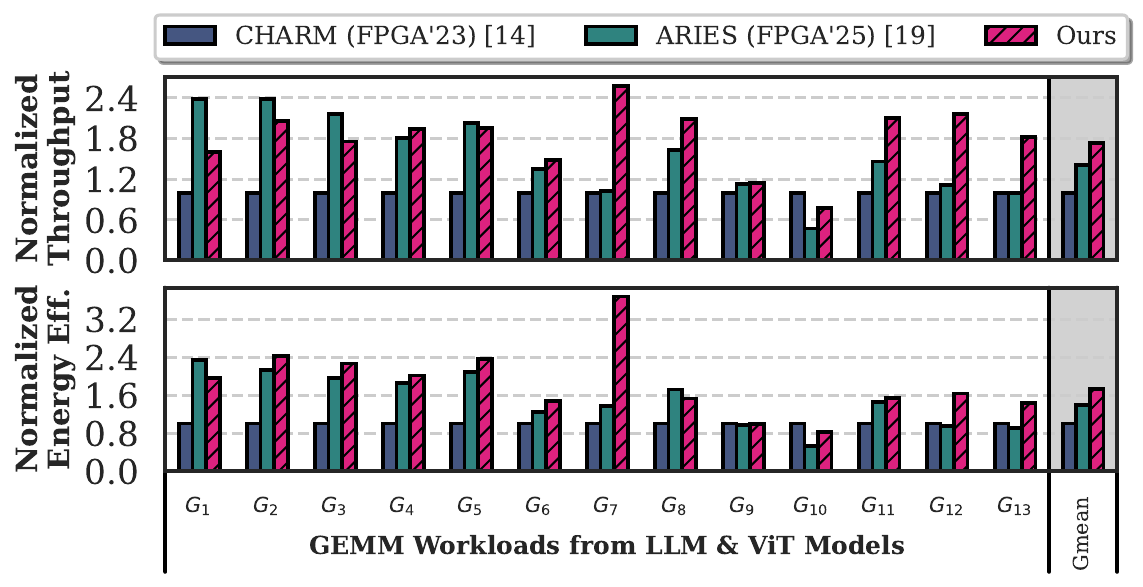}
    \caption{Throughput and energy efficiency of GEMM workloads ($G_n$) on Versal ACAP, normalized to CHARM and ordered by increasing arithmetic intensity.}
    \label{fig:throughput_energy_eff_prior_works}
\end{figure}

\subsection{GEMM Performance and Efficiency Evaluation}
We evaluate our framework in terms of performance and energy efficiency, using multiple GEMM workloads ($G_n$) found in popular LLM and ViT models, {incl. Swin-Tiny~\cite{liu2021swin}, DeiT-Base~\cite{touvron2021training}, Qwen2.5-0.5B~\cite{bai2023qwen}, LLaMA-3-1B~\cite{grattafiori2024llama}.}
These GEMM workloads are \emph{not} part of the training dataset, showcasing the ability of our approach to generalize to unseen workloads.
We compare our approach against state-of-the art approaches from CHARM~\cite{zhuang2023charm} and ARIES~\cite{zhuang2025aries} as well as embedded GPUs (Jetson).
Since no guidance for power consumption estimation is available from~\cite{zhuang2023charm,zhuang2025aries}, in both cases we use their highest throughput configuration.
For \emph{Ours} design, we use the generated hardware by our framework.
We report that the execution time of the DSE process using our ML model is minimal (less than 2~sec. per workload on an Intel Xeon Gold 6530), which is comparable to the execution time of DSE based on analytical equations from prior works.

\begin{figure}[t]
    \centering
    \includegraphics[width=1.0\linewidth]{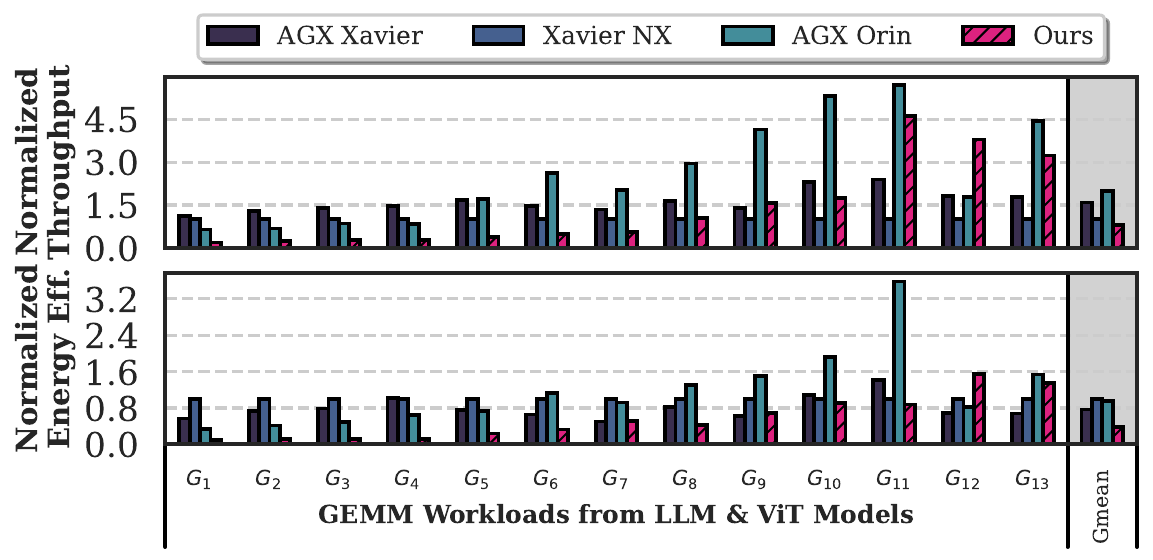}
    \caption{Throughput and energy efficiency of GEMM workloads ($G_n$) on Jetson GPUs, normalized to Xavier NX and ordered by increasing arithmetic intensity.}
    \label{fig:throughput_energy_eff_jetson}
\end{figure}

\subsubsection{Comparison with SoTA}
In Fig.~\ref{fig:throughput_energy_eff_prior_works} we list the normalized throughput and normalized energy efficiency of CHARM, ARIES and our framework.
We observe that our framework on average achieves higher throughput and energy efficiency compared to CHARM and ARIES.
Specifically, the geomean throughput speedup of \emph{Ours} compared to CHARM and ARIES are 1.73$\times$ and 1.23$\times$ respectively.
Compared to CHARM, our framework achieves throughput gains ranging from 0.78$\times$ to 2.58$\times$, while compared to ARIES, our framework achieves throughput gains ranging from 0.67$\times$ to 2.52$\times$.
When observing the normalized energy efficiency, the geomean gains of \emph{Ours} compared to CHARM and ARIES are 1.73$\times$ and 1.25$\times$ respectively.
Compared to CHARM, our framework achieves energy efficiency gains ranging from 0.82$\times$ to 3.68$\times$, while compared to ARIES, our framework achieves energy efficiency gains ranging from 0.84$\times$ to 2.69$\times$.
Additionally, \emph{Ours} is able to achieve higher energy efficiency even for workloads that lower throughput is achieved ($G_{2}, G_{3}$).
Overall, we observe that in the majority of workloads, our proposed framework is able to find more optimal GEMM mappings for both throughput and energy efficiency.

Table~\ref{tab:resource_utilization} shows the resource utilization of the generated designs for workloads $G_n$, with the best configuration per workload in bold.
For $G_2$–$G_7$, our framework achieves the highest energy efficiency, using on average 2.95$\times$ fewer AIEs than CHARM and ARIES.
Compared to our throughput-oriented designs, $G_4$–$G_6$ show up 24$\%$ lower AIE utilization on average.
PL usage is similar, with differences mainly in BRAM and URAM (e.g., up to 2.2$\times$ higher BRAM for $G_6$) to improve data reuse.
For throughput-oriented designs, our framework generally outperforms CHARM and ARIES using the same or fewer AIEs (except $G_1\dash G_3,G_8$) through better tiling that improves data reuse and DDR transfers.
Depending on the workload, our designs either use more BRAM/URAM (e.g., $G_{12}$) or achieve higher throughput with less (e.g., $G_4$) by allocating reuse buffers on the most beneficial GEMM dimension.
These results highlight the importance of accurate performance prediction to guide the DSE.

\begin{table}[t] 
\centering 
\caption{Resource Utilization by Workload $G_n$} 
\resizebox{\linewidth}{!}{%

\setlength{\tabcolsep}{3pt}

\begin{tabular}{cc*{16}{c}} 
\specialrule{.2em}{.0em}{0em} 
\rowcolor[HTML]{c7c3c3} 
\textbf{Target} & \textbf{Work} 
& \textbf{$G_1$} & \textbf{$G_2$} & \textbf{$G_3$} & \textbf{$G_4$} 
& \textbf{$G_5$} & \textbf{$G_6$} & \textbf{$G_7$} & \textbf{$G_8$} 
& \textbf{$G_9$} & \textbf{$G_{10}$} & \textbf{$G_{11}$} & \textbf{$G_{12}$} 
& \textbf{$G_{13}$} \\ 
\specialrule{.2em}{.0em}{0em} 

\multirow{4}{*}{\textbf{\#AIE}} 
& CHARM~\cite{zhuang2023charm} & 112 & 112 & 128 & 112 & 256 & 128 & 256 & 224 & \textbf{256} & \textbf{256} & 256 & 256 & 256 \\ 
& ARIES~\cite{zhuang2025aries} & \textbf{63} & \textbf{98} & \textbf{98} & 42 & \textbf{192} & 64 & 48 & \textbf{84} & 256 & 224 & 128 & 256 & 256 \\ \cline{2-15}
& \textbf{Ours (Throughput)} & 3 & 14 & 7 & \textbf{63} & 96 & \textbf{64} & \textbf{48} & \textbf{336} & \textbf{256} & 224 & \textbf{256} & \textbf{256} & \textbf{256} \\ 
& \textbf{Ours (Energy Eff.)} & 3 & \textbf{14} & \textbf{7} & \textbf{42} & \textbf{48} & \textbf{56} & \textbf{48} & 126 & \textbf{256} & 224 & \textbf{256} & \textbf{256} & \textbf{256} \\ \hline 

\multirow{4}{*}{\textbf{\makecell{BRAM \\ (\%)}}} 
& CHARM~\cite{zhuang2023charm} & 34.4 & 34.4 & 37.8 & 34.4 & 57.6 & 37.8 & 69.2 & 52.6 & \textbf{80.8} & \textbf{54.3} & 57.6 & 54.3 & 57.6 \\ 
& ARIES~\cite{zhuang2025aries} & \textbf{2.4} & \textbf{2.4} & \textbf{2.4} & 5.3 & \textbf{12.3} & 2.4 & 22.2 & \textbf{8.2} & 55.3 & 25.5 & 40.5 & 55.3 & 55.3 \\ \cline{2-15}
& \textbf{Ours (Throughput)} & 2.4 & 2.4 & 5.3 & \textbf{2.4} & 12.3 & \textbf{9.0} & \textbf{14.0} & \textbf{2.4} & \textbf{55.3} & 37.1 & \textbf{65.3} & \textbf{65.3} & \textbf{65.3} \\ 
& \textbf{Ours (Energy Eff.)} & 2.4 & \textbf{2.4} & \textbf{5.3} & \textbf{2.4} & \textbf{7.3} & \textbf{19.8} & \textbf{14.0} & 2.4 & \textbf{55.3} & 37.1 & \textbf{65.3} & \textbf{65.3} & \textbf{65.3} \\ \hline 

\multirow{4}{*}{\textbf{\makecell{URAM \\ (\%)}}} 
& CHARM~\cite{zhuang2023charm} & 0.0 & 0.0 & 0.0 & 0.0 & 0.0 & 0.0 & 0.0 & 0.0 & \textbf{0.0} & \textbf{55.3} & 27.7 & 27.7 & 27.7 \\ 
& ARIES~\cite{zhuang2025aries} & \textbf{9.1} & \textbf{21.2} & \textbf{6.0} & 18.1 & \textbf{27.6} & 6.9 & 2.6 & \textbf{18.1} & 27.6 & 13.8 & 13.8 & 13.8 & 13.8 \\ \cline{2-15}
& \textbf{Ours (Throughput)} & 1.3 & 3.0 & 3.0 & \textbf{9.1} & 13.8 & \textbf{6.9} & \textbf{5.2} & \textbf{36.3} & \textbf{13.8} & 13.8 & \textbf{27.6} & \textbf{27.6} & \textbf{27.6} \\ 
& \textbf{Ours (Energy Eff.)} & 1.3 & \textbf{3.0} & \textbf{3.0} & \textbf{6.0} & \textbf{6.9} & \textbf{3.5} & \textbf{5.2} & 18.1 & \textbf{13.8} & 13.8 & \textbf{27.6} & \textbf{27.6} & \textbf{27.6} \\ \hline 

\multirow{4}{*}{\textbf{\makecell{LUT \\ (\%)}}} 
& CHARM~\cite{zhuang2023charm} & 4.3 & 4.3 & 3.8 & 4.3 & 5.9 & 4.1 & 5.2 & 6.1 & \textbf{6.1} & \textbf{6.4} & 5.1 & 6.3 & 5.1 \\ 
& ARIES~\cite{zhuang2025aries} & \textbf{6.4} & \textbf{9.3} & \textbf{8.4} & 7.4 & \textbf{13.5} & 5.4 & 7.3 & \textbf{8.7} & 15.6 & 13.3 & 10.6 & 15.3 & 15.3 \\ \cline{2-15}
& \textbf{Ours (Throughput)} & 1.4 & 2.8 & 2.7 & \textbf{6.1} & 7.9 & \textbf{5.8} & \textbf{5.3} & \textbf{17.0} & \textbf{14.6} & 13.9 & \textbf{16.1} & \textbf{15.7} & \textbf{15.6} \\ 
& \textbf{Ours (Energy Eff.)} & 1.4 & \textbf{2.8} & \textbf{2.7} & \textbf{5.0} & \textbf{5.1} & \textbf{6.3} & \textbf{5.3} & 9.2 & \textbf{14.6} & 13.9 & \textbf{16.1} & \textbf{15.7} & \textbf{15.6} \\ \hline 

\multirow{4}{*}{\textbf{\makecell{FF \\ (\%)}}} 
& CHARM~\cite{zhuang2023charm} & 3.9 & 3.9 & 3.4 & 3.9 & 4.8 & 3.7 & 4.4 & 5.4 & \textbf{4.9} & \textbf{5.1} & 4.2 & 5.0 & 4.2 \\ 
& ARIES~\cite{zhuang2025aries} & \textbf{7.6} & \textbf{11.0} & \textbf{7.9} & 8.6 & \textbf{15.5} & 7.1 & 5.6 & \textbf{10.1} & 17.9 & 13.7 & 12.5 & 15.1 & 15.1 \\ \cline{2-15} 
& \textbf{Ours (Throughput)} & 1.5 & 2.9 & 2.7 & \textbf{7.8} & 8.9 & \textbf{7.3} & \textbf{6.3} & \textbf{20.5} & \textbf{12.9} & 13.9 & \textbf{17.9} & \textbf{17.8} & \textbf{17.8} \\ 
& \textbf{Ours (Energy Eff.)} & 1.5 & \textbf{2.9} & \textbf{2.7} & \textbf{6.4} & \textbf{6.3} & \textbf{5.9} & \textbf{6.3} & 11.1 & \textbf{12.9} & 13.9 & \textbf{17.9} & \textbf{17.8} & \textbf{17.8} \\ \hline 
2
\multirow{4}{*}{\textbf{\makecell{DSP \\ (\%)}}} 
& CHARM~\cite{zhuang2023charm} & 4.8 & 4.8 & 3.6 & 4.8 & 6.1 & 4.4 & 4.4 & 6.3 & \textbf{6.1} & \textbf{9.1} & 3.6 & 6.1 & 3.6 \\ 
& ARIES~\cite{zhuang2025aries} & \textbf{4.4} & \textbf{10.2} & \textbf{2.9} & 8.7 & \textbf{13.3} & 3.3 & 1.2 & \textbf{8.7} & 13.3 & 7.5 & 6.6 & 6.6 & 6.6 \\ \cline{2-15}
& \textbf{Ours (Throughput)} & 0.6 & 1.5 & 1.5 & \textbf{4.4} & 6.6 & \textbf{3.3} & \textbf{2.5} & \textbf{17.4} & \textbf{6.6} & 7.7 & \textbf{13.3} & \textbf{13.3} & \textbf{13.3} \\ 
& \textbf{Ours (Energy Eff.)} & 0.6 & \textbf{1.5} & \textbf{1.5} & \textbf{2.9} & \textbf{3.3} & \textbf{1.7} & \textbf{2.5} & 8.7 & \textbf{6.6} & 7.7 & \textbf{13.3} & \textbf{13.3} & \textbf{13.3} \\ \hline 
\specialrule{.2em}{.0em}{0em} 
\end{tabular}%
} 
\label{tab:resource_utilization} 
\end{table} 

\subsubsection{Comparison with GPUs}
From Fig.~\ref{fig:throughput_energy_eff_jetson} we observe that on average NVIDIA's Jetson outperform VCK190 in energy efficiency and throughput.
However, different behavior is observed for the more memory- from the more compute-bound workloads.
The larger gains over VCK190 are observed for the smaller workloads $G_{1}\dash G_{8}$, which is expected due to the large difference in memory bandwidth, ranging from 2.33-8$\times$ as shown in Table~\ref{table:exp_setup}.
When moving to the more compute-bound workloads $G_{9}\dash G_{13}$ we observe the performance gap decreasing, with VCK190 closely following or outperforming AGX Xavier and Xavier NX in throughput and energy efficiency.
Notably, for $G_{12}$ VCK190 outperforms AGX Orin in throughput and energy efficiency by 2.3$\times$ and 2$\times$ respectively.

\subsubsection{Quality of Generated Pareto Fronts}
\begin{figure}
    \centering
    \includegraphics[width=1.0\linewidth]{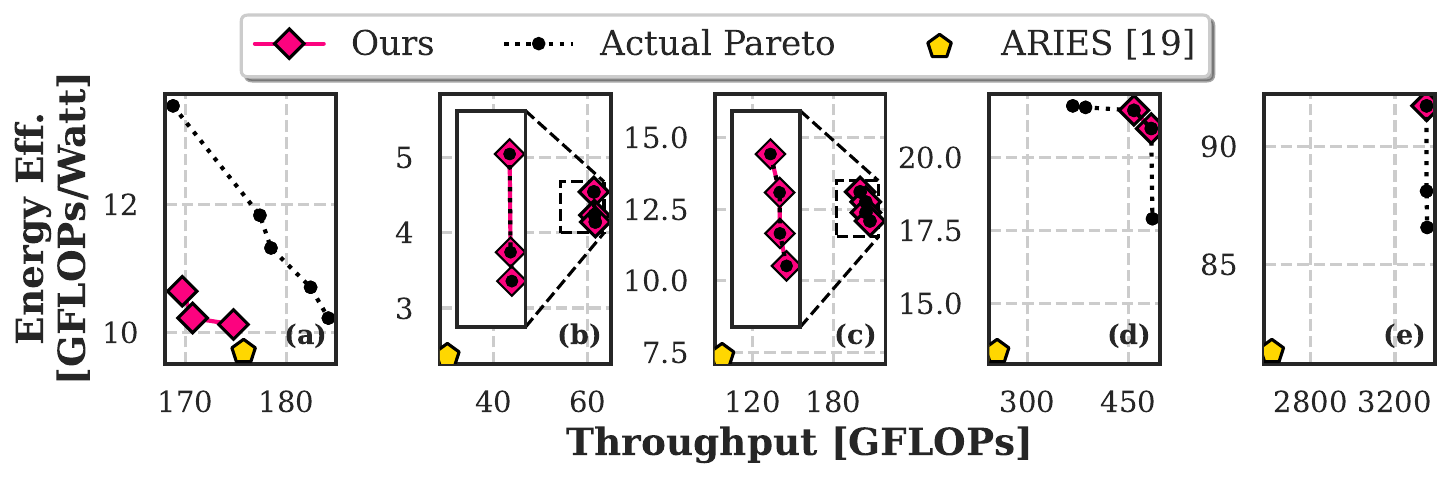}
    \caption{Pareto solutions generated by: \emph{i)} ARIES~\cite{zhuang2025aries}, \emph{ii)}~Our framework. Black points indicate actual Pareto-front.}
    \label{fig:eval:pareto}
\end{figure}

Fig.~\ref{fig:eval:pareto} shows the Pareto-optimal solutions for 5 GEMM workloads generated by: \emph{i)} ARIES (yellow) and \emph{ii)} our framework (pink), compared against the actual Pareto front from exhaustive experiments (black).
Our framework closely matches the true Pareto front for most workloads (b)–(e), with predicted points lying directly on it, matching even in number of points on the Pareto-front for (b) and (c). 
In workload (a), where it does not fully align with the Pareto-front, our framework still surpasses prior work.
Overall, our framework achieves a 2.18$\times$ higher hypervolume area on geomean, with gains up to 3.84$\times$. 

\section{Conclusion}


In this work we present an automatic framework for Versal ACAP that finds optimal GEMM hardware mappings for either energy efficiency or performance, using a machine learning model to predict performance and power consumption.
Unlike prior analytical approaches, it uses a machine learning model to predict performance and power more accurately, achieving average 1.2-1.3$\times$ improvements in both throughput and energy efficiency over state-of-the-art methods.

\bibliographystyle{ieeetr}
\bibliography{refs/refs}

\end{document}